\documentclass[12pt,preprint]{aastex}
\usepackage{times}

\newcommand{\dtsn}{$\delta\tau_\mathrm{SN}$}
\newcommand{\ms}{m~s$^{-1}$}

\shorttitle{Profile of Solar Interior Meridional Flow}
\shortauthors{Zhao et al.}

\begin{document}
\title{Detection of Equatorward Meridional Flow and Evidence of Double-Cell 
       Meridional Circulation inside the Sun}

\author{Junwei Zhao\altaffilmark{1}, R.~S.~Bogart\altaffilmark{1},
A.~G.~Kosovichev\altaffilmark{1}, T.~L.~Duvall, Jr.\altaffilmark{2},
and Thomas Hartlep\altaffilmark{1}}
\altaffiltext{1}{W.~W.~Hansen Experimental Physics Laboratory, Stanford
University, Stanford, CA94305-4085}
\altaffiltext{2}{Solar Physics Laboratory, NASA Goddard Space Flight
Center, Greenbelt, MD20771}

\begin{abstract}
Meridional flow in the solar interior plays an important role in redistributing
angular momentum and transporting magnetic flux inside the Sun. Although 
it has long been recognized that the meridional flow is predominantly 
poleward at the Sun's surface and in its shallow interior, the location 
of the equatorward return flow and the meridional flow profile in the 
deeper interior remain unclear. Using the first two years of continuous 
helioseismology observations from the {\it Solar Dynamics Observatory} / 
Helioseismic Magnetic Imager, we analyze travel times of acoustic waves 
that propagate through different depths of the solar interior carrying 
information about the solar interior dynamics. After removing a systematic 
center-to-limb effect in the helioseismic measurements and performing 
inversions for flow speed, we find that the poleward meridional flow of 
a speed of 15~\ms\ extends in depth from the photosphere to about 
0.91~R$_\sun$. An equatorward flow of a speed of 10~\ms\ is found 
between 0.82 to 0.91~R$_\sun$ in the middle of the convection zone. Our 
analysis also shows evidence of that the meridional flow turns poleward again 
below 0.82~R$_\sun$, indicating an existence of a second meridional 
circulation cell below the shallower one. This double-cell meridional 
circulation profile with an equatorward flow shallower than previously 
thought suggests a rethinking of how magnetic field is generated and 
redistributed inside the Sun.
\end{abstract}


\section{Introduction}

Meridional flow inside the Sun plays an important role in redistributing 
rotational angular momentum and transporting magnetic flux \citep{wan89}, 
and is crucial to our understanding of the strength and duration of 
sunspot cycles \citep{hat10, dik10} according to flux-transport dynamo 
theories \citep{cho95, dik99}. At the Sun's surface and in its shallow 
interior to at least 30~Mm in depth, the meridional flow is predominantly 
poleward with a peak speed of approximately 20~\ms, as measured by 
tracer tracking \citep[e.g.,][]{kom93, sva07, hat10}, direct Doppler-shift 
observations \citep[e.g.,][]{ulr10}, and local helioseismology analyses
\citep[e.g.,][]{gil97, hab02, zha04, gon08}. The poleward plasma flow 
transports the surface magnetic flux from low latitudes to the polar region, 
causing the periodic reversals of the global magnetic field, a process
important to the prediction of the solar cycles \citep{dik10}. The speed and 
variability of the meridional flow also play an important role in determining 
the strength and duration of the solar cycles, and the unusually 
long activity minimum at the end of Solar Cycle 23 during 2007 -- 2010 
was thought to be associated with an increase of the meridional flow 
speed during the declining phase of the previous cycle \citep{hat10}.
Therefore, an accurate determination of the meridional flow profile is 
crucial to our understanding and prediction of solar magnetic activities.

Although the poleward meridional flow at the solar surface and in shallow 
depths has been well studied, the depth and speed profile of the equatorward 
return flow, which is expected to exist inside the solar convection zone 
to meet the mass conservation, largely remains a puzzle. It is generally 
assumed that the return flow is located near the base of the convection zone,
although no convincing evidence has been reported.  Earlier helioseismic 
analyses of {\it Solar Heliospheric Observatory} / Michelson Doppler Imager 
\citep[{\it SOHO}/MDI;][]{sch95} data were unable to directly detect 
an equatorward flow, but estimated that the return flow had a speed of 2~\ms\ 
near the bottom of the convection zone after applying a mass-conservation 
constraint \citep{gil99}. This single-cell circulation profile has 
been later widely employed in the flux-transport dynamo simulations and 
solar cycle prediction models \citep[e.g.,][]{dik10, jia07}. A later
helioseismic frequency-shift analysis put the equatorward flow at a much 
shallower depth of 40~Mm \citep{mit07}, but this was questioned
by \citet{gou10}. More recently, an analysis based on tracking of photospheric 
supergranulation gave an estimate of the return flow depth below 50~Mm 
\citep{hat12}, however, this analysis was unable to give an unambiguous 
profile of the meridional circulation as a function of both latitude and 
depth inside the Sun. A more robust analysis utilizing a well-established 
helioseismology technique, e.g., time-distance helioseismology \citep{duv93}, 
is needed to examine the deep meridional circulation profile.

The continuous Doppler observations by the Helioseismic and Magnetic Imager 
\citep[HMI;][]{sch12a, sch12b} onboard the recently launched {\it Solar 
Dynamics Observatory} mission \citep[{\it SDO};][]{pes12} allow us to
measure and detect the long-sought equatorward flow. Our analysis, which
takes into account the systematic center-to-limb effect that was recently
found in the local helioseismology analysis techniques \citep{zha12},
gives a two-dimensional cross-section picture of the meridional 
flow inside the nearly entire solar convection zone, and reveals a
double-cell circulation with the equatorward flow located near the
middle of the convection zone. In this Letter, we describe our analysis
procedure in Sec.~2, present measurements and inversion results in
Sec.~3, and then discuss these results in Sec.~4.

\section{Data Analysis}

The {\it SDO}/HMI provides continuous full-disk Doppler-shift observations 
of the Sun with a 45-sec cadence for helioseismology studies. Time-distance 
helioseismology \citep{duv93} measures travel times of acoustic waves, 
which propagate through the solar interior, by cross-correlating oscillation 
signals observed at different locations on the surface. The measured 
travel-time anomalies are usually interpreted in terms of wave-speed 
perturbations and mass flows along the wave path. This analysis technique 
has already provided rich information about the solar interior structures 
and dynamics \citep{giz10}. 

Determining the internal meridional flow is one important task in solar 
physics and yet one difficult challenge due to the low speed of the flow, 
which requires a careful analysis of potential systematic errors. In 
particular, recent studies have revealed a previously unreported systematic 
center-to-limb effect in the acoustic travel times measured by the 
time-distance technique \citep{zha12}. The systematic travel-time variation 
depends on the angular distance from the solar disk center, and exhibits 
different magnitude, sometimes even opposite signs, when measured using 
different HMI observables and using different measurement distances. It 
is not quite clear what causes this systematic effect, and a recent study 
suggested that it might be partially due to the interactions of acoustic 
waves with the vertical flows in solar granules \citep{bal12}. \citet{zha12} 
suggested that this systematic effect should be removed before inverting 
the measured acoustic travel times for interior meridional flow, and proposed 
to use the travel-time shifts measured in the east-west direction along 
the equator as proxies of the systematic effect. Their inversion results, 
after removal of this effect following their suggested method, showed a 
reduction of $\sim$10~\ms\ in the inverted meridional flow speed \citep{zha12}.
We believe that this systematic effect was responsible for many past 
helioseismology study failures to reliably detect the equatorward meridional 
flow, and that the removal of this effect will play an important role 
in studying the meridional flow in the solar interior.

\begin{figure}
\epsscale{0.95}
\plotone{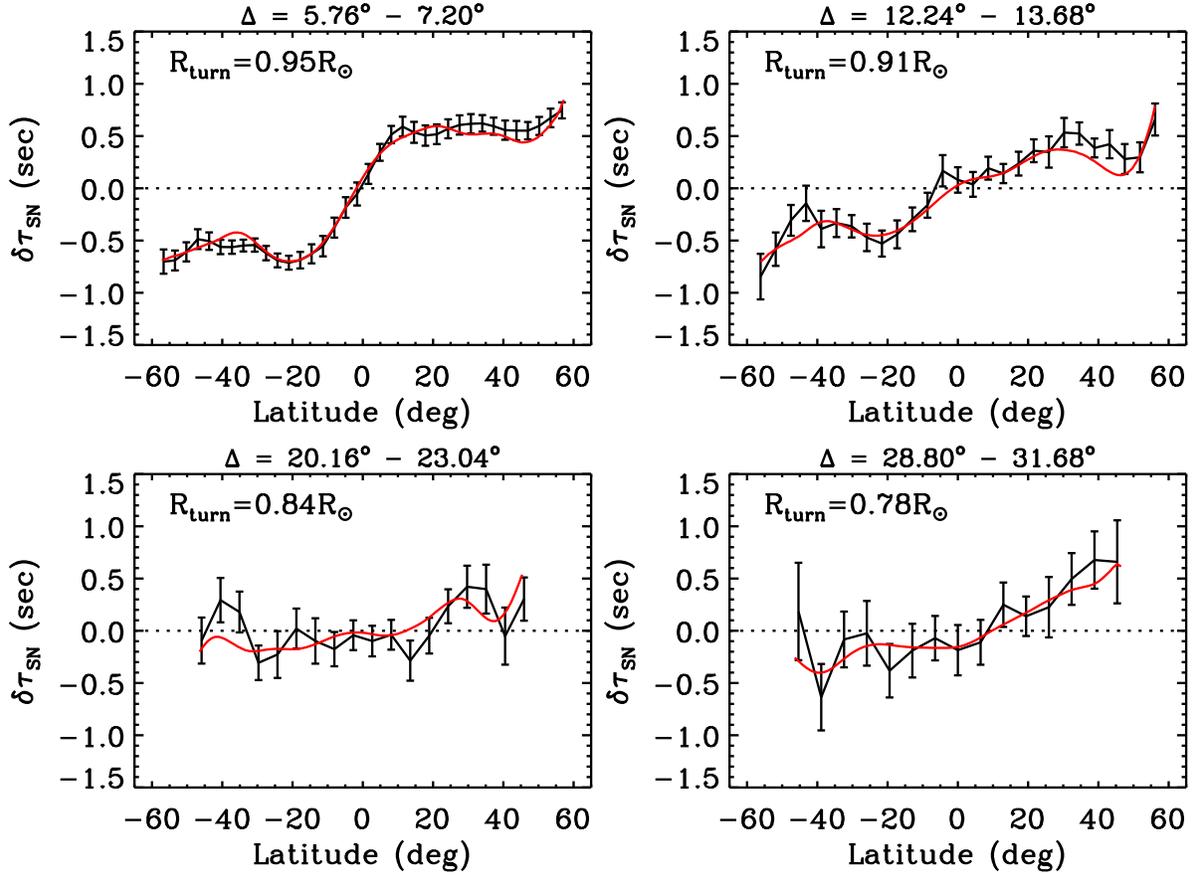}
\caption{Measured acoustic travel-time differences, \dtsn, shown as black 
curves for selected measurement distances. The measurement distances $\Delta$ 
and corresponding radii of acoustic wave lower turning points, 
R$_\mathrm{turn}$, which are derived from the ray-path approximation, are 
marked in each panel. Red curves are the \dtsn\ calculated from our 
inversion results displayed in Figure~\ref{merid_flow}, showing that 
our inversion results fit the measurements well. Error bars for the 
red curves are similar to those for the black curves.}
\label{time_lat}
\end{figure}

We use the HMI Doppler-shift data covering its first 2-year period 
from 2010 May 1 through 2012 April 30, during which the solar magnetic 
activity was generally low. For each observing day, we track the data 
with a uniform Carrington rotation rate, remap the tracked data onto 
the heliographic coordinate using Postel's projection with a sampling scale 
of 0$\fdg$18 pixel$^{-1}$, and then apply a running-difference filter to
remove solar convection and low-frequency signals. Note that no other 
filtering is applied in our analysis in order to avoid any complications 
in the Fourier domain when dealing with very large spatial scales. A 
deep-focusing time-distance measurement scheme (shown in Figure~\ref{scheme}), 
which was described and validated by \citet{har13}, is utilized. For each 
measurement, the solar oscillation signals are first averaged inside two 
30$\degr$-long single-pixel-wide concentric arcs, located opposite to each 
other in the north and south directions, and then cross-correlated to 
extract acoustic wave packets traveling between these arcs along the 
wave paths through the solar interior. In order to make a robust 
center-to-limb effect correction, we limit the selection of oscillation 
signals within 15$\degr$ from the central meridian. The center-to-limb 
variations are estimated from the east-west travel-time measurements along 
an equatorial band following the same geometry as used in the north-south 
measurements along the meridian band. The same procedure is repeated for 
each of the 60 measurement distances ranging from 2$\fdg$16 to 44$\fdg$64, 
covering a radial range from approximately 0.7~R$_{\sun}$ to the surface 
according to the acoustic ray theory. The cross-correlation functions are 
averaged for the same latitudes, and then averaged again over one-month 
intervals. The Gabor wavelet fitting \citep{kos97} is performed to derive 
the acoustic travel times for all the measurement distances. For each distance, 
the travel times are averaged again over the whole 2-year period to 
obtain the final results. The standard errors after the systematic effect
correction for each measurement distance, estimated from the 24 measurement 
periods, are displayed as error bars in Figures~\ref{time_lat} and 
\ref{time_dep}. The errors of the center-to-limb effect follow a similar 
pattern as in \dtsn.

\section{Results}
\subsection{Results of Travel-Time Measurements}

Figure~\ref{time_lat} shows some selected examples of \dtsn, which is 
defined as the travel-time differences between the southward and northward 
propagating acoustic waves after corrected for the systematic center-to-limb 
variation, as a function of latitude. A small systematic offset caused 
by the imperfect alignment of the HMI instrument relative to the solar 
rotation axis is also removed following the same approach by \citet{gil97} 
and \citet{hat10}. Examples of \dtsn\ as functions of the measurement 
distance, or radius of the lower turning point of acoustic wave paths 
estimated from acoustic ray-path approximation, for selected latitudes 
are shown in Figure~\ref{time_dep}a. The \dtsn, caused by the internal 
meridional flows, are of an order of 1~sec for short distances (corresponding 
to the wave paths confined to the shallow interior and reflecting the 
shallow poleward flow), and decrease rapidly with the measurement distance 
(or the turning-point radius). The values of \dtsn\ drop close to zero 
and even change sign for certain latitudes, and then become to increase 
again. This trend is essentially the same for both hemispheres, although 
there are clear differences between the two. 

To gain more confidence in these measurements and the procedure correcting 
the systematic effect, we also analyze the Doppler-shift data acquired by 
{\it SOHO}/MDI \citep{sch95}. MDI observed the Sun from 1996 through 2011, 
and its Dynamics Program, usually lasting two months each year, had 
continuous full-disk Doppler-shift observations for helioseismic studies. 
To avoid complications caused by solar magnetic activity and for the purpose 
of a better comparison with the HMI results obtained during a relatively 
quiet period, we select the MDI Dynamics Program data from the solar 
minimum years of 1996 -- 1998 and 2006 -- 2010, and perform the analysis 
following the exactly same procedure as we do for HMI data. Note that 
although the full-disk data from HMI and MDI have different spatial 
resolutions, both data are remapped to a lower resolution of 0$\fdg$18 
pixel$^{-1}$ before being used for the helioseismic analysis. We find 
that despite the differences in the observed spectral lines and different 
observing periods, the \dtsn\ from the two instruments are in reasonable 
agreement after the systematic effect correction (Figure~\ref{time_dep}a). 
It is worth pointing out that the original measurements of \dtsn\ from 
both instruments differ substantially before the removal of the 
center-to-limb effect (Figure~\ref{time_dep}b). This indicates that 
the center-to-limb variations from the HMI and MDI are significantly 
different, which is believed due to that they observe different spectral 
lines, and also shows that our procedure to remove the systematic effect 
is reasonable and robust.

\subsection{Results from Inversion}

We then invert the corrected \dtsn\ from HMI for the interior meridional 
flow speed as a function of latitude and depth. Ray-approximation 
sensitivity kernels are utilized in the inversion, and they are prepared
following the same geometry and averaging procedures as in the measuring. 
Three-dimensional kernels are first computed in spherical coordinates,
and then collapsed into one meridional plane. Basically, the sensitivity 
kernels show thin curves of the ray path, with the strongest sensitivity
located near the lower turning points.
We linearize the equation that links the travel-time variations with 
the interior flow field: $\delta \tau_\mathrm{diff} = -2 \int_\Gamma 
(\mathbf{n \cdot v})/c^2 \mathrm{d} s$, where $\Gamma$ is acoustic ray 
path, $\mathbf{v}$ is flow velocity, $c$ is sound speed, and $s$ is 
distance along the ray path \citep{kos96}. The linearized equations are 
then solved in a sense of least squares using the LSQR algorithm, an 
iterative linear equation solver \citep{pai82}. In our inversions, the 
radial component of the flow is not taken into account but it is not 
expected that this will affect our inversion results significantly 
because the flows are predominantly horizontal near the deep-focus points 
where our measurement is more sensitive. However, an inversion procedure 
including the radial flow will be developed in our future studies.

We carry out two sets of inversions with and without the mass-conservation 
constraint. For the results inverted from only the helioseismic measurements 
without applying the mass-conservation constraint, a two-dimensional 
cross-section view of the meridional flow velocity and flow profiles 
at some selected depths and latitudes are displayed in Figure~\ref{merid_flow}. 
The \dtsn\ calculated from a forward procedure, which utilizes the same 
equation as given above that links travel time and interior flows, using 
the inverted velocity are overplotted in Figures~\ref{time_lat} and 
\ref{time_dep}a to compare with the measured \dtsn. It can be found 
that they are in good agreement, demonstrating that our inversion results
fit the measurement quite well. Basically, our 
inversion results show that the poleward flow of 15~\ms\ extends from the 
surface to a depth of approximately 65~Mm, i.e., about 0.91~R$_{\sun}$. 
Below this, the flow direction turns equatorward with a maximum speed of 
about 10~\ms. Both the depth and the thickness of the zone of return 
flow are clearly latitude dependent. The equatorward flow shows a 
significant hemispheric asymmetry, stronger in the northern hemisphere 
and weaker in the southern. Below about 0.82~R$_{\sun}$, the flow again 
becomes poleward, also with a substantial hemispheric asymmetry. 
Errors from the inversion are typically 1~\ms\ near the surface, and 
are as large as 10~\ms\ below 0.80~R$_{\sun}$. Our inversion stops 
at about 0.75~R$_{\sun}$ because the noise prevents reliable inferences 
in the deeper layers. Longer observations are required to increase 
the signal-to-noise ratio.

Another set of inversion with the mass-conservation constraint (results are 
not shown in a figure) demonstrates that the basic flow profile remains 
similar to that without applying the mass constraint. The major 
difference is that the zone of equatorward flow becomes thicker, and 
the poleward flow below 0.82~R$_{\sun}$ becomes slower.

\begin{figure}
\epsscale{0.75}
\plotone{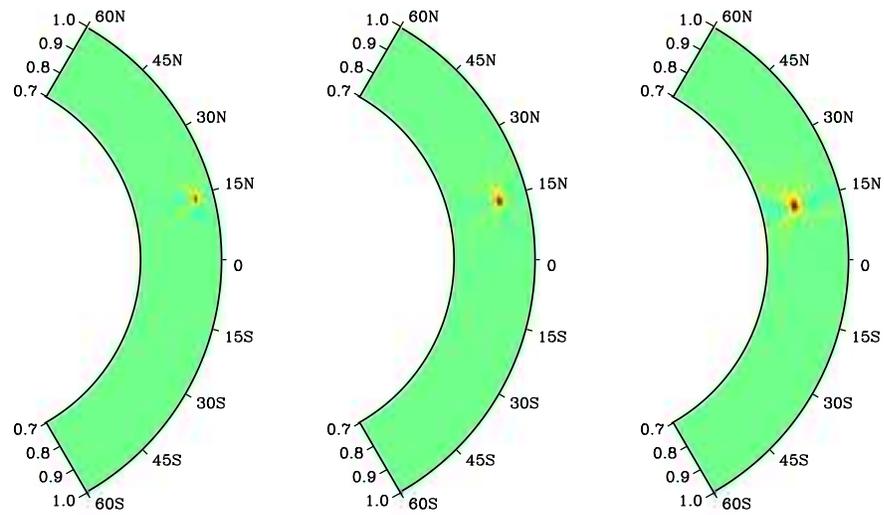}
%\caption{(a) Same as Figure~\ref{merid_flow}a, but for the inversion results
%with the mass-conservation constraint. (b) Same as Figure~\ref{time_dep}a but 
%comparing the HMI measurements with the \dtsn\ calculated from the inversion 
%results of both with and without the mass-conservation constraint. Panels (c), 
%(d), and (e) are averaging kernels obtained from our inversion procedure when 
%the target is located at 15$\degr$N and depths of 50, 75, and 125 Mm, 
%respectively. }
\caption{Averaging kernels obtained from the inversion procedure when the 
target is located at the latitude of 15$\degr$N and depths of 50, 75, and 
125 Mm, respectively.}
\label{mass_kernel}
\end{figure}

The averaging kernels from our inversions for some selected target depths are 
displayed in Figure~\ref{mass_kernel}. It can be seen that the kernel is 
well localized in the shallow areas, and becomes broader with the increase 
of depth, but are still localized in the latitudinal direction. The 
negative side lobes above and below the targeted area in the radial 
direction also become stronger in amplitude and larger in size with the 
depth. The averaging kernels remain the same for the same depth but different 
latitudes.

%Inversion results with the mass-conservation constraint are shown in 
%Figure~\ref{mass_kernel}a; Figure~\ref{mass_kernel}b shows a comparison
%of the measured \dtsn\ with the \dtsn\ calculated from the forward procedure 
%using velocity profiles obtained from the both sets of inversions with
%and without the mass constraint. Basically, the meridional flow profile 
%from the inversion with the mass constraint shows a thicker zone of 
%equatorward flow, a slightly faster equatorward flow speed, and a slower 
%poleward flow beneath this zone. Compared with the measured data, the 
%calculated \dtsn\ is in worse agreement than that from the inversion without 
%the mass constraint. However, we also acknowledge that the mass conservation 
%used in our inversion may not provide an accurate constraint because the 
%mass and flow below 0.75~R$_{\sun}$ and below the convection zone are 
%not able to be included in our inversion.

\section{Discussion}

\begin{figure}
\epsscale{0.80}
\plotone{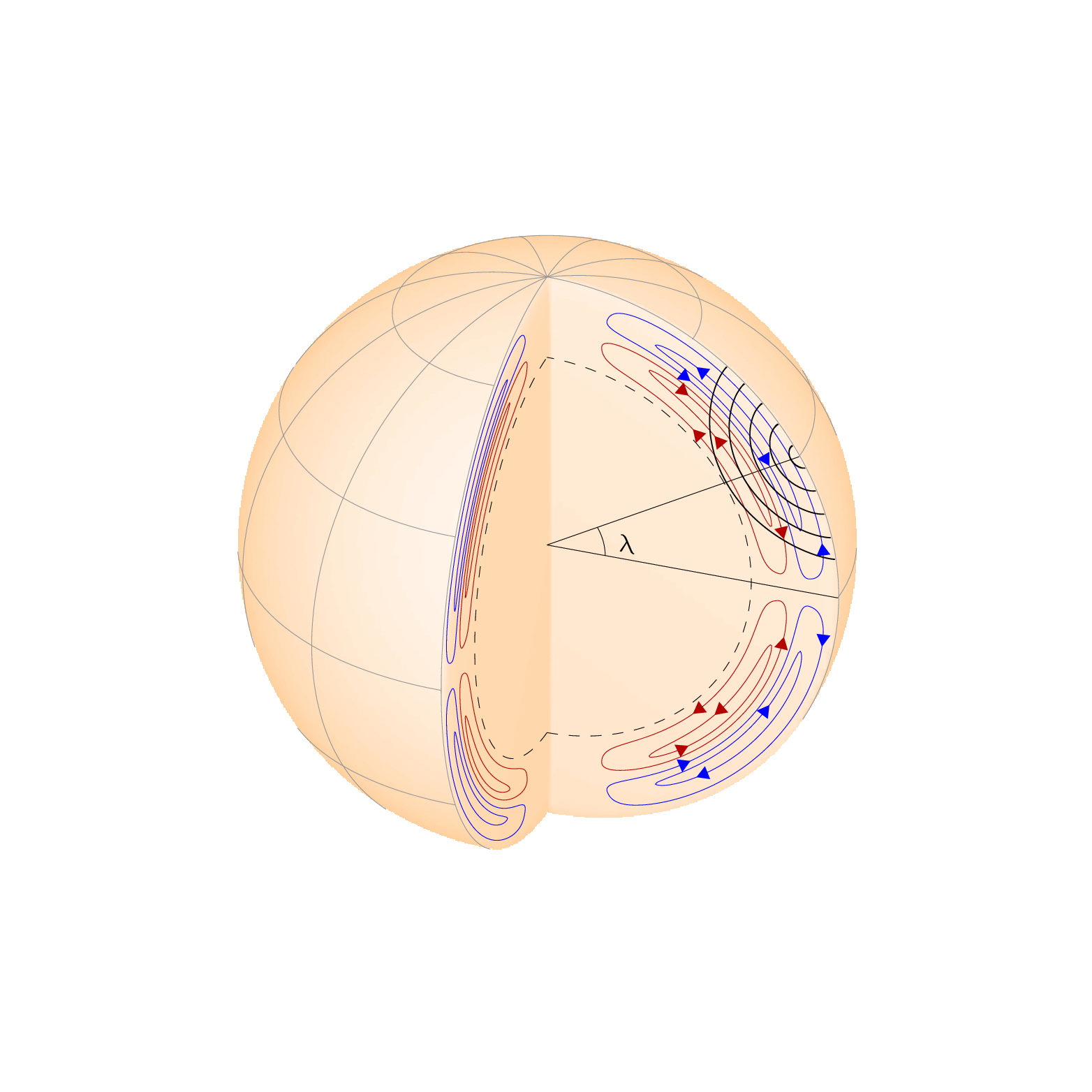}
\caption{Diagram of the deep-focusing time-distance measurement scheme, 
with black curves showing some samples of acoustic wave paths. Blue and 
red streamlines show a schematic structure of the meridional circulation 
illustrating our results. Black dashed lines show the bottom of the 
convection zone at 0.7~R$_{\sun}$.}
\label{scheme}
\end{figure}

Our analysis reveals a new picture of the Sun's interior meridional 
circulation, with poleward flows from the surface to 0.91~R$_{\sun}$ 
and from 0.82~R$_{\sun}$ to at least 0.75~R$_{\sun}$, and an equatorward 
flow located between these in a layer of about 0.09~R$_{\sun}$ thick. 
This seems to form two meridional circulation cells, one beneath the 
other in the radial direction (Figure~\ref{scheme}). Not being able to 
invert the flow profile beneath 0.75~R$_{\sun}$ and above the latitude 
of 60$\degr$, we cannot exclude the possibility of more circulation 
cells in both radial and latitudinal directions. %Our inversion results 
%with and without the mass-conservation constraint differ in terms of 
%flow speed and thickness of the zone of the equatorward flow, but in 
%general the picture of the double-cell structure with an equatorward 
%flow located in the middle of the convection zone basically remains 
%the same.

\begin{figure}
\epsscale{0.60}
\plotone{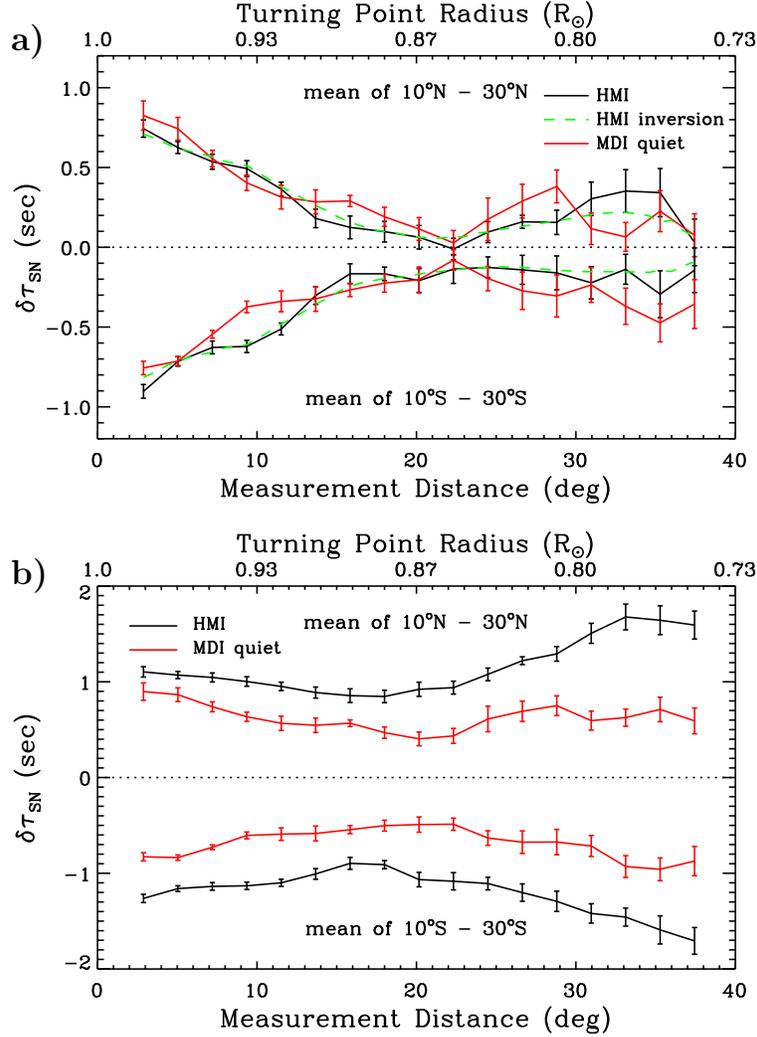}
\caption{(a) Comparison of the measured acoustic travel-time differences, 
\dtsn, from the HMI and MDI data after removal of the systematic center-to-limb
variations, together with the \dtsn\ calculated from the interior velocity 
inverted from the HMI measurements. Error bars for the green dashed curves
are similar to those for the black curves.
(b) Comparison of the \dtsn\ from the HMI and MDI quiet-period observations 
before the removal of the systematic effect.}
\label{time_dep}
\end{figure}

\begin{figure}
\epsscale{0.95}
\plotone{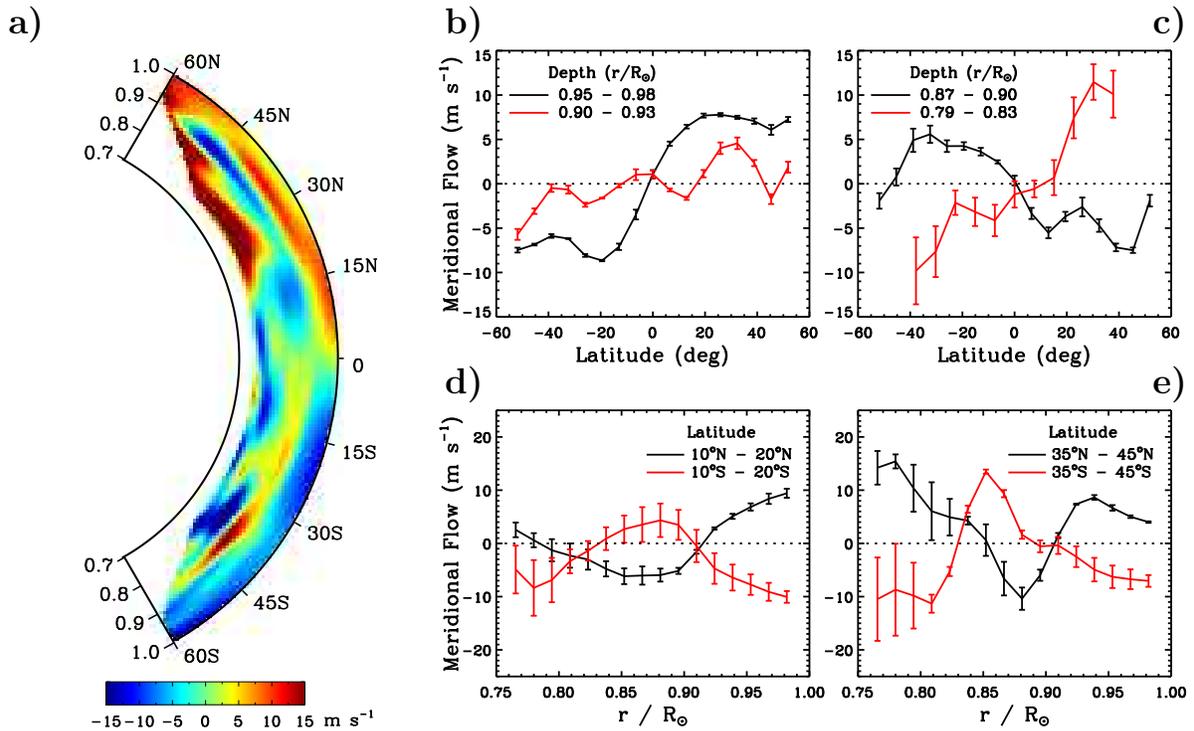}
\caption{Meridional flow profile, obtained by inverting the measured acoustic
travel times. Panel (a) shows a cross-section view of the meridional-flow 
profile, with the positive velocity directing northward. Panels (b) and (c) 
show the inverted velocity as functions of latitude averaged over several depth 
intervals. Panels (d) and (e) show the velocity as functions of depth 
averaged over different latitudinal bands.}
\label{merid_flow}
\end{figure}

The interior meridional flow profile reported in this Letter is obtained 
after the removal of the systematic center-to-limb effect, therefore, 
a precise determination of this flow profile relies largely on a precise 
determination of that systematic effect. We follow the empirical approach 
of using the east-west travel-time measurements along the equator to 
represent this systematic effect, however, without knowing the exact 
cause of this effect (as already pointed out in Sec.~2), a small deviation 
of the real systematic effect from this representation can result in 
a significant change of the inverted meridional flow profile shown in 
Figure~\ref{merid_flow}. However, the comparison of MDI and HMI results,
which carry different amount of systematic effects, gives us confidence 
that our effect correction procedure is reasonable (see Figure~\ref{time_dep}).
Still, we believe it is necessary to develop a different systematic-effect 
correction procedure in the future, after the cause of the center-to-limb 
effect is better known, to give a more robust inference of the deep meridional 
flow profile.

Based on a single-cell circulation model with the equatorward flow, peaking
at 3~\ms, located near the bottom of the convection zone, \citet{bra08}
estimated that more than one decade of continuous observations is required 
for a reliable detection of the return flow, which is expected to cause
an acoustic travel-time shift of about 0.01~sec. However, the meridional
flow profile determined from our measurements differs from the model used 
in their estimation. Furthermore, Figure~\ref{time_dep} shows that the 
travel-time shift caused by the shallow poleward flow is partly offset by 
the equatorward flow in the middle of the convection zone, leaving the flow 
near the bottom easier to measure. This explains why two years of
helioseismic observations is able to probe the deep interior meridional 
flow. Certainly, longer observations will enable us to reliably 
invert the flow beneath 0.75~R$_{\sun}$ and enhance the S/N ratio 
through the entire convection zone, and the capability of the 
shorter-time measurement allows us to study the temporal evolution 
of the deep meridional flow with lower S/N ratio.

This new picture of the solar interior meridional circulation differs 
substantially from the previously widely-believed picture of a single-cell 
circulation with the equatorward flow near the bottom of the 
convection zone. Through removing a systematic center-to-limb effect that
was only recently identified, our analysis corrects and improves the 
previous solar interior meridional flow profile given by \citet{gil99} 
using a similar analysis procedure. The new meridional circulation 
profile poses a challenge to the flux-transport dynamo models \citep{cha10}, 
but provides more physical constraints to these models creating a 
new opportunity to further understand how magnetic field is generated 
and how magnetic flux is transported inside the Sun. Past dynamo 
simulations have already demonstrated that a meridional circulation 
profile with multiple cells might not be able to reproduce the butterfly 
diagram and the phase relationship between the toroidal and poloidal 
fields as observed, unless the dynamo model was reconsidered \citep{jou07}. 
However, on the other hand, solar convection simulations have shown 
the possibility of multi-cell circulation with a shallow equatorward flow 
\citep[e.g.,][]{mie06, kap12, gue13}, demonstrating that our analysis 
results are reasonable. Moreover, a recent dynamo simulation, with the 
double-cell meridional circulation profile incorporated, showed that 
the solar magnetic properties could be robustly reproduced after taking 
into consideration of turbulent pumping, turbulent diffusivity, and other 
factors \citep{pip13}. All these studies, together with our observational 
results, suggest a rethinking of how the solar magnetic flux 
is generated and transported inside the Sun.

%The profile of the solar meridional flow does not usually stay unchanged
%through different phases of one solar cycle. With more HMI observations
%accumulating, we will continue to monitor how the interior meridional
%flow profile evolves with the solar cycle.

\acknowledgments
We thank the two anonymous referees whose comments help to improve the 
presentation and quality of this Letter. SDO is a NASA mission, and HMI 
project is supported by NASA contract NAS5-02139 to Stanford University.

\newpage

\newpage

\newpage

\newpage

\newpage

\end{document}